\documentstyle[prd,aps]{revtex}
\begin{document}
\twocolumn[\hsize\textwidth\columnwidth\hsize\csname
@twocolumnfalse\endcsname
\draft
\title{Trans-Planckian Particle Creation in Cosmology and
Ultra-High Energy Cosmic Rays}
\author{Alexei A. Starobinsky$^1$ and Igor I. Tkachev$^2$}
\address{$^1$ Landau Institute for Theoretical Physics RAS,
Kosygina 2, Moscow, 119334, Russia}
\address{$^2$ Theory Division, CERN, CH-1211 Geneva 23, Switzerland}
\date{\today}
\maketitle
\begin{abstract}
We consider observational constraints on creation of particles
induced by hypothetical trans-Planckian effects during
the current stage of the Universe expansion.
We show that compatibility with the diffuse $\gamma$-ray
background measured by the EGRET experiment strongly restricts
this creation. In particular, it rules out the possibility to
detect signatures of such short distance effects in anisotropies
of the cosmic microwave background radiation. On the other hand, 
a possibility that some part of the ultra-high energy cosmic rays 
originates from new trans-Planckian physics remains open.
\end{abstract}
\vskip2pc]

\vspace{1cm}

Recently, much interest was attracted to
the study of possible deviations of the dispersion law of quantum
ultrarelativistic particles from the standard one $\omega(k)=k$
at very large ("trans-Planckian") momenta $k>M$ (presumably,
$M\sim M_{\rm Pl}=\sqrt G$; we put $\hbar=c=1$ in this paper). Such a
suggestion was previously discussed in quantum theory of black
holes \cite{Unruh} (where it does not lead to any new observable
effects), but then it was applied in cosmology \cite{MB}. Reasons
for the existence of such an effect may follow from explicit
breaking of the Lorentz invariance either induced by the existence
of additional spatial dimensions (e.g., with "asymmetric warping" 
of usual 4D curved space-time~\cite{Csaki}), or suggested by 
analogy with quasiparticles in quantum liquids \cite{Vol}. 
Non-standard dispersion laws arise in non-commutative geometry~
\cite{Chu} and $\kappa$-Poincare symmetry algebra \cite{Kow}, too.

Almost all attempts to find observational signatures of this
effect in cosmology were related to its influence on spectra
of scalar perturbations and gravitational waves generated during
inflation. However, as was emphasized in \cite{St01}, if any
correction to these spectra arises at all, it means creation
of real particles with ultra-high energies (caused by some new
trans-Planckian physics) due to {\em any} expansion of the Universe.
In particular, it should occur at the present time, too. Note, that
there is even no qualitative difference between the type of the
Universe expansion during a de Sitter (inflationary) stage in the
early Universe and nowadays: both are accelerating ones. Of course,
the present value $H_0$ of the Hubble parameter $H\equiv \dot a/a$,
where $a(t)$ is the scale factor of the Friedmann-Robertson-Walker
(FRW) cosmological model and $H_0$ is the Hubble constant, is much
less than $H$ during inflation. But, as we will see, it is much
easier to detect particles with ultra-high energy created now than
those created long time ago during inflation (in spite of the fact
discussed below that the number of created particles is second
order in the parameter of non-adiabaticity $|\beta(k)|$ while
corrections to the spectra of inflationary perturbations are first
order in $|\beta(k)|$).

Following the general approach of \cite{St01} (see also more
recent papers \cite{Dan}), we will phenomenologically describe
the effect of ultra-high energy particle creation in cosmology
due to unknown trans-Planckian physics in the following way.
Expansion of the Universe results in redshifting of spatial
momenta: $k=n/a(t),~n=const$ where $k=|{\bf k}|$ (in the case
of a non-commutative geometry, the quantities which are
redshifted $\propto a^{-1}$ are not exactly the usual momenta
${\bf k}$, but the difference between them and ${\bf k}$ becomes
small for $k\ll M$, see \cite{Chu}). As a result, wave equations
for time-dependent parts of quantum field operators in the
Heisenberg representation have the following form in the regime of
large momenta $k\gg H$:
\begin{equation}
\ddot\phi_k + 3H\dot\phi_k + \omega^2 \left({n\over a}\right)
\phi_k=0
\label{scalar}
\end{equation}
for scalar particles, and
\begin{equation}
A''_k +a^2\omega^2 \left({n\over a}\right)A_k = 0
\label{vector}
\end{equation}
where the dot denotes the derivative with respect to the time $t$,
the prime - the derivative with respect to the conformal
time $\eta = \int dt/a(t)$, $A_k$ is some quantity
characterizing the electromagnetic field (it is proportional to
covariant components of the vector-potential ${\bf A}$ in the
standard case), and the 3D spatial Fourier expansion is assumed.
Note that, in principle, $\omega(k)$
for the electromagnetic field may depend on photon polarization,
too. Deviation of $\omega(k)$ from the standard law $\omega = k$
for $k\stackrel {>}{\sim} M$ results in breaking of conformal
invariance for photons (and massless neutrinos, too), so photon
creation in the FRW metric becomes possible. Below we will argue
that massive particles with a restmass $m\ll M$ must be created
as well (even if $m\gg H$), if creation of massless particles
is not suppressed.

Let $H\ll M$. Then generic solutions of Eqs.
(\ref{scalar},\ref{vector}) have the following form in the WKB
regime $H\ll k \ll M$ (in the leading WKB approximation):
\begin{eqnarray}
\phi_k= {1\over \sqrt{2\omega a^3}}\left(\alpha_{n,0}
e^{-i\int \omega dt} + \beta_{n,0} e^{i\int \omega dt}\right)~,
\label{scalWKB} \\
A_k= {1\over \sqrt{2\omega a}}\left(\alpha_{n,1}
e^{-i\int \omega dt} + \beta_{n,1} e^{i\int \omega dt}\right)~,
\label{vecWKB} \\
|\alpha_{n,s}|^2 - |\beta_{n,s}|^2=1~,~~s=0,1~
\label{norm}
\end{eqnarray}
(we omit the spin index $s$ further).

Usually, the adiabatic vacuum $\beta_n = 0$ is assumed for all
modes of all quantum fields. However, trans-Planckian physics may
result in a nonzero $\beta_n$ (its actual value may be different
for quantum fields of different spins and even for different
polarizations, but we will not consider the latter possibility).
So, supposing that particles with $k\gg M$ do not exist as
individual particles or are not observable for other reasons
(since we don't see them after all), we arrive to the following
observational picture of the effect in consideration: in the
course of the Universe expansion, pairs of particles and
antiparticles with super-high energy $M~(\sim M_{\rm Pl})$ are
spontaneously created at the moment when their momentum
$k(t)\equiv n/a(t)=M$, and their occupation number
is $|\beta_n|^2$. The corresponding correction coefficient
${\cal K}^2(n)$ to the power spectrum of inflationary perturbations
is obtained by matching of Eq. (\ref{scalWKB}) (or its analog for
gravitational waves) to the exact solution of massless scalar wave
equation in the (approximately) de Sitter background with the
Hubble parameter $H$ estimated at the moment of the first Hubble
radius crossing $k(t)=H$. It is equal to:
\begin{equation}
{\cal K}^2(n) = |\alpha_n - \beta_n|^2~.
\label{corr}
\end{equation}
We will see below that $|\beta_n|$ should be small. Then $\alpha_n$
can be made unity by a phase rotation, and ${\cal K}^2(n)= 1 -
2\,{\rm Re}\,\beta_n$. Its difference from unity is first order in
$|\beta_n|$.

Our approach is to take $\alpha_n$ and $\beta_n$ (subjected to the
condition (\ref{norm})) as phenomenological quantities which
should finally follow from a concrete model of non-trivial
trans-Planckian physics, and investigate how they are limited by
present observational data. Thus, we consider real
particle creation (corresponding to an imaginary part of the
effective action of quantum fields in a FRW background) only.
This should be contrasted to real, vacuum polarization
corrections to the effective action considered, e.g., in
\cite{Am}. The latter corrections result in a refraction
index different from unity for radiation. They can be strongly
limited by observations of distant $\gamma$-bursts \cite{grbM}.
Note also that corrections to the effective volume in phase
space leading to "trans-Planckian damping" which were recently
proposed in \cite{Has} (in particular, they may explain why
particles with $k\gg M$ are not observable) can be easily
incorporated in the formalism used by changing the overall
time-dependent prefactors in Eqs. (\ref{scalWKB},\ref{vecWKB}).

In \cite{St01}, the first step in this investigation was made by
considering back reaction of created ultra-high energy gravitons
on the Universe expansion at present. It was assumed that
$\beta_n$ has the following expansion in terms of the small
parameter $H_n/M$ where $H_n\equiv H(t_n)$ is the Hubble
parameter estimated at the moment of the trans-Planckian border
crossing $n=Ma(t_n)$ for each Fourier field mode ${\bf k}$:
\begin{equation}
\beta_n = \beta_n^{(0)} + \beta_n^{(1)}{H_n\over M}
+ ...
\label{exp}
\end{equation}
Then it was shown that the first term in (\ref{exp}) is very
suppressed: $|\beta_n^{(0)}|^2 \stackrel{<}{\sim}  H_0^2M_{\rm  Pl}^2/M^4 = 
10^{-122} \, M_{\rm Pl}^4 / M^4$,
while the
second term is bounded by $|\beta_n^{(1)}|^2 \ll M_{\rm Pl}^2/M^2$
(so, it is also suppressed if $M\sim M_{\rm Pl}$). Note that time
translation invariance (which we don't want to abandon) requires
$|\beta_n^{(0)}|^2$ and $|\beta_n^{(1)}|^2$ to be independent on $n$
that was noted in \cite{St01}. On the other hand, the phase
of $\beta_n$ is $n$-dependent and may be large. This leads
to oscillations in ${\cal K}^2(n)$ and in resulting inflationary
perturbation spectra which, however, are unobservable for
$H\ll M$ due to their high frequency in $k$-space~\cite{Dan,Shan}.

The first, $H$-independent term in (\ref{exp}) describes "pure"
trans-Planckian particle creation where the Universe expansion
plays a kinematic role only. The second term in (\ref{exp})
is responsible for a mixed effect where both small-scale
trans-Planckian physics and large-scale space-time curvature
participate. A concrete toy model producing the latter term
was proposed in \cite{St01}, namely, the quantum state of any
Fourier field mode ${\bf k}$ which has a minimal energy density
just at the moment of the trans-Planckian border crossing (this
state differs from the adiabatic vacuum in the next
term of the WKB expansion). Since the minimal energy state may
not appear as a result of the adiabatic evolution in the WKB
regime $|\dot\omega|\ll \omega^2$ (even for a non-standard
dispersion law), this model implicitly assumes that something
radical happens for $k>M$: either that any mode does not exist
in this regime at all, and is instantaneously "created" at the
moment when its momentum falls down to $M$, or that the WKB
condition is suddenly violated for $k>M$, i.e., because of
$\omega(k)$ becoming very small for $k>M$ (as it occurs, e.g.,
in the model considered in~\cite{Mers}). Then, if
$\omega(k)=k$ for $k<M$ exactly, the model leads to
$|\beta_n^{(1)}|=1/2$ for minimally coupled scalar
particles~(\ref{scalar}) (see \cite{St01}, the recent papers
\cite{Dan} arrived to essentially the same result).

To create photons, some deviation from the standard
dispersion law $\omega(k)=k$ should exist for $k\le M$
already. Let us assume that the quantity to be diagonalized
for each Fourier mode ${\bf k}$ is $\tilde\varepsilon_k =
(\hat A'^2_k+a^2\omega^2\hat A^2_k)/2 a^4$,
then equations for $\alpha_n$ and $\beta_n$ in the
representation (\ref{vecWKB}) take the form (c.f. \cite{ZS}
for the case of a conformally coupled massive scalar field):
\begin{equation}
\alpha'_n = {\Omega'_n\over 2\Omega_n}~e^{2i\int \Omega_n
d\eta}~\beta_n~, ~~~\Omega_n = a\omega\left({n\over a}\right)
={n\omega \over k}~,
\label{aphoton}
\end{equation}
\begin{equation}
\beta'_n = {\Omega'_n\over 2\Omega_n}~
e^{-2i\int \Omega_n d\eta}~\alpha_n~.
\label{bphoton}
\end{equation}
The diagonalization condition at $\eta=\eta_0(n)$ (when $k=M$)
is $\beta_n(\eta_0)=0$. If particle creation is small, $|\beta_n|
\ll 1$, then $\beta_n\approx -i\,(\Omega'_n/\Omega_n^2)_{\eta_0}/4$
(up to a phase factor and an additional strongly oscillating
term). Therefore,
\begin{equation}
|\beta_n^{(1)}|= {M\over 4}\left[ {k^2\over \omega^2}
\left|{d\over dk}\left({\omega\over k}\right)\right|\right]_
{k=M}\sim 1
\label{photon}
\end{equation}
for photons.

Note that the expression (\ref{photon}) remains valid for
conformally coupled massive particles as far as their restmass
$m\ll M$. So, this toy model shows that the second term in the
expansion (\ref{exp}) need not be suppressed for massive
particles with $m\gg H$. This remarkable fact may be
understood using the following argument: any non-standard
dispersion law $\omega(k)$ is equivalent to the appearance of an
effective mass term $m^2(k)\equiv \omega^2(k)-k^2$ ($m^2$ may
be negative, of course). For $k\sim M$, where a significant
deviation from the standard dispersion law occurs, the rest mass
$m^2(0)$ is completely irrelevant.

Equations for creation of massive fermions in a FRW background
are similar to those in the case of conformally coupled massive
scalar particles (with an additional multiplier $n/ma$ in
the r.h.s. of Eqs.~(\ref{aphoton},\ref{bphoton}) for the
standard dispersion law $\omega^2=k^2+m^2$, see e.g.
\cite{St84}). Therefore, if photons are created due to
trans-Planckian effects at all, one may expect that massive
fermions with $m\ll M$ including leptons are created with a
comparable (or even slightly larger) rate due to the present
expansion of the Universe.

Now we make a next step and study limits on trans-Planckian
particle creation following from the direct observability of
created particles (photons, in particular). Also, we omit the
assumption $M\sim M_{\rm Pl}$ and consider the case $M\ll
M_{\rm Pl}$, too. We show that high energy cosmic rays data
require much more suppression of $\beta_n^{(0)}$ and
$\beta_n^{(1)}$ as compared to the results obtained in
\cite{St01}.


The measured flux of ultra-high energy cosmic rays (UHECR) extends
to energies of order $E \sim E_0 \equiv 10^{11}$ GeV only. On the
other hand, a typical energy of particles emerging from the
trans-Planckian region can be much higher, up to $E \sim 10^{19}$
GeV. May the highest energy particles pass undetected ? The answer
is negative. First, measurements place the following constraint
on the integral flux of high energy particles (see, e.g.,
\cite{uhecr}):
\begin{equation}
F_{E > E_0} \approx 10^{-2} \; {\rm km}^{-2}\, {\rm yr}^{-1}\,
{\rm sr}^{-1}
\approx 10^{-71}\;  {\rm GeV}^3 \, {\rm sr}^{-1}\; .
\label{eq:uhecr_flux}
\end{equation}
Second, the Universe is not transparent for high energy radiation.
Particles which are injected with any $E>E_0$ will rapidly (on
the cosmological time scale) migrate into a lower energy range. For
our purposes, it it sufficient to consider attenuation of high
energy particles on photons of the cosmic microwave background
(CMB) radiation.

Protons loose energy in the process of pion photoproduction. This
gives rise to the famous Greisen-Zatsepin-Kuzmin (GZK) cut-off. The
attenuation length for this process (that is the distance over which
the energy of a primary particle decreases by one e-fold) is less than
$20$ Mpc at $E > E_0$.  Roughly half of released energy ends up in the
electromagnetic cascade, the rest is carried out by neutrinos. The
Universe becomes transparent for protons with $E \approx E_0$.
Therefore, the number of protons which could have been produced by
trans-Planckian effects (and which conserve) is subject to the
constraint (\ref{eq:uhecr_flux}). This can be re-written as a
constraint on the quantum gravity scale $M$ in a way similar to what
follows.  However, a somewhat stronger and less model dependent
constraint can be obtained by considering an electromagnetic cascade
which migrates to even lower energies. From this point of view, it is
unimportant whether the electromagnetic cascade was initiated by
propagation of high energy protons, or photons (or, to this end,
electrons) which were created by the trans-Planckian effects
directly. Even neutrino production in the trans-Planckian
region is not harmless. Neutrino will create the electromagnetic
cascade in interactions with the cosmic background of relic
neutrinos. Since about 1\% of high energy neutrinos interact over
the horizon scale \cite{weiler}, our final constraint, Eq.
(\ref{eq:constraint1}), would be only an order of magnitude weaker
even in the unrealistic case of pure neutrino creation. For these
reasons, we concentrate on the constraint imposed by the 
electromagnetic cascade in what follows.

A high energy photon cascades to lower energies in the chain of
the following reactions.  First, it creates $e^+ e^-$ pairs in
collisions with CMB photons. Secondary electrons re-create
photons with energies somewhat lower than the energy of the
original photon via the inverse Compton process, and so on. The
corresponding attenuation length at $E \gg E_0$ is about 0.1 of
the present horizon size, and it is even smaller for smaller
energies. Therefore, the cascade migrates to lower energies until
it reaches the sub-TeV scale which corresponds to the threshold
of pair creation on cosmic backgrounds.

Therefore, the integrated energy flux of particles emerging from
the trans-Planckian region may not exceed the integrated energy
flux in the sub-TeV range where the diffuse $\gamma$-ray
background was measured by the EGRET telescope \cite{Egret}.
The measured value of this background is
\begin{equation}
S_0 \approx 10^3 \; {\rm eV\, cm^{-2}\, s^{-1}\, sr^{-1} }
\sim 10^{-58}\; {\rm GeV}^4 \, {\rm sr}^{-1}\; .
\label{eq:egret}
\end{equation}

Let us relate this flux to the energy production rate. The rate
of growth of energy density in particles emerging from the
trans-Planckian region due to the expansion of the Universe is
\cite{St01}
\begin{equation}
J \equiv \frac{d (a^4 \epsilon) }{a^4 dt} = {g N M^4H\over 2\pi^2}
|\beta_n|^2~.
\label{eq:flux}
\end{equation}
In this relation, both particles and antiparticles are counted,
$g=2$ for photons and neutrinos, $g=4$ for massive fermions. 
N counts for all particle species which can create the
electromagnetic cascade at the end, since one expects that the 
trans-Plankian creation is ``democratic'' and insensitive to
particle masses as far as $m\ll M$. Omitting neutrino, $N = 26$ 
in the standard model. In supersymmetric or Grand Unified 
models,~$N \sim 10^2 - 10^3$. The
integrated flux of energy accumulated during the age of the
Universe will be $S_1 \approx J\, H^{-1}$. Requiring $S_1 < S_0$,
we get
\begin{equation}
|\beta^{(0)}_{n}|^2 < 10^{-133} \;\; \frac{1}{N} \; \left(
\frac{M_{\rm Pl}}{M}\right)^4~.
\label{eq:constraint}
\end{equation}
We see that the constraint on the $\beta_n^{(0)}$ term in the
decomposition (\ref{exp}) is very strong. Thus, this term should
be practically absent regardless of the value of $M$. A
contribution from the second term is strongly suppressed by the
small quantity $H_0^2/M_{\rm Pl}^2 \approx 10^{-122}$. As a
result, for the $\beta_n^{(1)}$ coefficient we obtain
\begin{equation}
|\beta_n^{(1)}| <  10^{-6} \;\; \frac{1}{\sqrt{N}} \; \frac{M_
{\rm Pl}}{M} \; .
\label{eq:constraint1}
\end{equation}


In recent literature (see e.g. \cite{BM02}) there were optimistic
expectations regarding possible imprints of short distance physics
on the spectrum of CMB anisotropies generated in the inflationary
scenario of the early Universe. Let us estimate now the impact of
the restriction (\ref{eq:constraint1}) on a possible magnitude of
the effect. According to Eqs. (\ref{corr}), (\ref{exp}),
(\ref{eq:constraint}), a fractional correction to the power
spectrum of inflationary perturbations which arise due to
trans-Planckian physics is given by
\begin{equation}
\frac{\delta P}{P} = \beta_n^{(1)} \; \frac{H_{\rm inf}}{M}
\label{corr2}
\end{equation}
where $H_{\rm inf}$ is the value of the Hubble parameter during
the last $60$ e-folds of inflation, $H_{\rm inf}/M_{\rm Pl} <
10^{-5}$. In view of the constraint (\ref{eq:constraint1}), we
find
\begin{equation}
\frac{\delta P}{P} < 10^{-11} \;\; \frac{1}{\sqrt{N}}\; \left(\frac
{M_{\rm Pl}}{M}\right)^2\; .
\label{corr3}
\end{equation}
On the other hand, astrophysical data on the constancy of the
light velocity yield the lower limit $M > 10^{15}$ GeV
\cite{grbM}.\footnote{Strictly speaking, this limit was obtained 
assuming that a correction to the standard dispersion law for 
$k\to 0$ starts with the cubic term, $\omega^2 = k^2(1 \pm (k/M) + 
\dots)$. If the cubic term is absent and the correction begins from 
a larger power of $k/M$, there is no lower limit on $M$. However, the 
constraint (\ref{eq:constraint1}) remains valid. So, even in this 
specific case, to obtain significant corrections to the perturbation 
power spectrum generated during inflation, either a specific mechanism 
for trans-Planckian particle creation producing $|\beta_n^{(1)}|\gg 1$ 
should be invented, or one has to postulate a low $M\le 10^{-6}\; 
M_{\rm Pl}$ which is not compatible with the condition $H_{inf}\ll M$ 
(necessary for general relativistic description of inflation and 
generation of perturbations) for many inflationary models.}
This gives ${\delta P}/{P} < 10^{-3}$ for the
maximum possible magnitude of corrections to the perturbation
power spectrum. We conclude that trans-Planckian particle
creation is so strongly restricted by observations of UHECR
that it will be impossible to detect signatures of short distance
physics in CMB anisotropies, since the allowed contribution
is smaller than the cosmic variance at all multipoles
of interest, $l < 10^4$.

Returning to UHECR themselves, one may consider a speculative
possibility that observed events above the GZK cut-off energy
are due to peculiarities of trans-Planckian physics. However,
trans-Planckian creation of particles would occur homogeneously
in the Universe, and therefore should lead to the GZK cut-off 
in the spectrum of created protons at high energies and to the pile-up 
of protons at $E \sim 4\times 10^{19}$ eV. Thus, protons can not 
explain super-GZK events despite the trans-Planckian creation does occur 
within the GZK sphere of $\sim 50$ Mpc, from where protons can reach us. 
On the other hand the attenuation length for photons grows
with energy and therefore photons may produce spectrum of cosmic
rays compatible with the AGASA data \cite{AGASA} at highest energies.
One problem which may arise here is related to an overall
normalization. At $E \sim 10^{20}$ 
eV the attenuation length for photons is about 100 times smaller
than the horizon scale. This gives the distance scale to sources
which contribute to the flux at ultra-high energies.
On the other hand, by-products of the 
electromagnetic cascade will pile-up at the EGRET energies and are 
accumulated from the entire Universe. On this grounds, one expects 
that the ratio of the energy flux in UHECR 
($S \sim 10^{-60}\; {\rm Gev^4\;  sr^{-1}}$, 
see Eq. (\ref{eq:uhecr_flux})) to the diffuse EGRET 
background can not be larger than $0.01$. This value 
comfortably fits the data, and the numerical coincidence
may indicate that these two backgrounds can be related indeed.  
However, to maintain this level of the UHECR flux in photons
one should assume small extragalactic magnetic fields and small
universal radio background (c.f. \cite{photons}). 
In addition, one would need to fine-tune 
the rate of trans-Planckian creation to the level of the observed
UHECR flux. Also, this mechanism is  disfavoured by the observed
angular clustering of UHECR \cite{AGASA,clustering}. One should note,
however, that the same problems arise in many other models which 
attempt to explain super-GZK events.

We conclude that at least some part of cosmic rays with energies
beyond the GZK limit may have origin due to new physics in the 
trans-Planckian region. This striking possibility remains open and 
deserves further study, while the constraint (\ref{corr3}) makes 
expected contribution of trans-Planckian physics to the CMB 
anisotropies to be unobservable.

The authors thank CITA, University of Toronto, where this project
was started, for hospitality. A.S. was also partially supported
by RFBR, grant 02-02-16817, and by the RAS Research Programme
"Quantum Macrophysics".

\end{document}